\documentclass[twocolumn,showpacs,preprintnumbers,amsmath,amssymb]{revtex4}

\usepackage{graphicx}
\usepackage{bm}
\usepackage{color}
\usepackage{amsmath}
\usepackage{amsfonts}
\usepackage{amssymb}
\usepackage{epstopdf}

\begin{document}

\preprint{APS/123-QED}

\title{Quasi-two-dimensional Bose\,--\,Einstein condensation of lattice bosons in spin-1/2 XXZ ferromagnet K$_2$CuF$_4$}

\author{Satoshi Hirata}
\author{Nobuyuki Kurita}
\author{Motoki Yamada}
\author{Hidekazu Tanaka} 
\affiliation{Department of Physics, Tokyo Institute of Technology, Meguro-ku, Tokyo 152-8551, Japan
}
\date{\today}

\begin{abstract}
K$_2$CuF$_4$ is magnetically described as a spin-1/2, quasi-two-dimensional (2D), square-lattice XXZ ferromagnet with weak easy-plane anisotropy. The magnetic ordering for an applied magnetic field $H$ parallel to the $c$ axis is equivalent to the Bose\,--\,Einstein condensation (BEC) of lattice bosons, as discussed by Matsubara and Matsuda [Prog. Theor. Phys.  \textbf{16}, 569 (1956)]. Magnetization and specific heat measurements were performed to obtain the temperature vs magnetic field phase diagram for $H\,{\parallel}\,c$. The phase boundary between polarized and ordered phases was found to be expressed by the power law $H_{\rm c}(T)\,{-}\,H_{\rm c}(0)\,{\propto}\,T^\phi$ with exponent $\phi\,{\approx}\,1.0$ in a wide temperature range, in agreement with the theory of quasi-2D BEC.
\end{abstract}


\maketitle
\section{Introduction} 
The Bose\,--\,Einstein condensation (BEC) of magnetic quasi-particles in gapped quantum magnets has been attracting considerable attention from the viewpoint of the quantum phase transition (QPT)\,\cite{Nikuni,Rice,Giamarchi1,Zapf_rev}. Dimerized quantum magnets, such as TlCuCl$_3$\,\cite{Oosawa1,Rueegg}, KCuCl$_3$\,\cite{Oosawa2}, BaCuSi$_2$O$_6$\,\cite{Jaime,Sebastian}, Ba$_2$CuSi$_2$O$_6$Cl$_2$\,\cite{Okada} and Ba$_3$Mn$_2$O$_8$\,\cite{Uchida,Samulon}, and quantum magnets with strong uniaxial easy-plane single-ion anisotropy, such as NiCl$_2$-4SC(NH$_2$)$_2$ (abbreviated to DTN)\,\cite{Zapf,Yin} and CsFeCl$_3$\,\cite{Kurita}, undergo transverse magnetic ordering when subjected to an external magnetic field. Because the triplet or doublet components with $S^z\,{=}\,{+}\,1$ are regarded as lattice bosons, these systems can be mapped onto a system of interacting bosons. Thus, the magnetic-field-induced transverse ordering can be understood as the BEC of the lattice bosons. At $T\,{=}\,0$ K, these systems undergo a QPT at a critical field $H_{\rm c}(0)$ with varying magnetic field. This critical field corresponds to the quantum critical point (QCP), which separates the quantum paramagnetic state and BEC state. The low-temperature magnetic properties of the gapped quantum magnets have successfully been described by the BEC theory rather than the mean field theory of the spin system\,\cite{TY,TM}. 

In the vicinity of the QCP, the phase boundary between the paramagnetic and ordered phases is described by the power law
\begin{eqnarray}
H_{\rm c}(T)-H_{\rm c}(0)\propto T^{\phi},
\label{eq:power}
\end{eqnarray}
where $H_{\rm c}(T)$ is the critical field at temperature $T$ and ${\phi}$ is the critical exponent\,\cite{Nikuni,Giamarchi1,Zapf_rev}. In a three-dimensional (3D) system, the critical exponent is given by ${\phi}_{\rm BEC}\,{=}\,3/2$~\cite{Nikuni,Giamarchi1,Zapf_rev}, which has been confirmed for several gapped quantum magnets, such as TlCuCl$_3$\,\cite{YamadaF1,YamadaF2}, DTN\,\cite{Zapf,Yin} and (CH$_3$)$_2$CHNH$_3$CuCl$_3$\,\cite{Tsujii}. On the other hand, for a quasi-2D system, the theory\,\cite{Syromyatnikov} predicts that the phase boundary is described with ${\phi}\,{\approx}\,1.0$ in a wide temperature range except at sufficiently low temperatures that are lower than the magnitude of the interlayer interaction. However, this prediction has not been sufficiently verified experimentally. To explore the quasi-2D BEC of lattice bosons and its critical behavior, we performed magnetization and specific heat measurements on K$_2$CuF$_4$, which is described as a spin-1/2, quasi-2D, square-lattice XXZ ferromagnet with weak easy-plane anisotropy\,\cite{Knox,Hidaka,Legrand,YamadaI,Khomskii,Funahashi,Yamazaki,Kubo,Hirakawa3,Sasaki,Hirakawa,Hirakawa2}.

In 1956, using the correspondence between spin operators and boson operators, i.e., $S_i^+\,{\leftrightarrow}\,a_i^{\dagger}$ and $S_i^-\,{\leftrightarrow}\,a_i$, Matsubara and Matsuda\,\cite{Matsubara} demonstrated that the partition function of a spin-1/2 XXZ ferromagnet in a magnetic field, which is expressed as
\begin{eqnarray}
{\cal H}_{\rm F}=\hspace{-1mm}&-&\hspace{-1mm}\sum_{\langle i,j\rangle}\left\{J^{\perp}\left(S_i^xS_j^x+S_i^yS_j^y\right)+J^{\parallel}S_i^zS_j^z\right\}\nonumber\\
\hspace{-1mm}&-&\hspace{-1mm}g{\mu}_{\rm B}H\sum_{i}S_i^z,
\label{eq:Spin_model}
\end{eqnarray}
is mathematically identical to the grand partition function of a system of lattice bosons with a hard core and the nearest-neighbor attractive interaction, which is represented as
\begin{eqnarray}
{\cal H}_{\rm L}=\frac{{\hbar}^2}{2md^2}\sum_{\langle i,j\rangle}(a_i^{\dagger}-a_j^{\dagger})(a_i-a_j)-v_0\sum_{\langle i,j\rangle}a_i^{\dagger}a_ia_j^{\dagger}a_j,
\label{eq:Lattice_boson}
\end{eqnarray}
where $m$ is the mass of a boson, $d$ is the lattice spacing and $v_0\,{>}\,0$ is the interaction constant. These particle quantities are related to the exchange constants as ${\hbar}^2/md^2\,{=}\,J^{\perp}$ and $v_0\,{=}\,J^{\parallel}$. The magnetic field parallel to the $z$ direction corresponds to the chemical potential of the boson system, i.e., ${\mu}\,{=}\,g{\mu}_{\rm B}H\,{+}\,z(J^{\perp}\,{-}\,J^{\parallel})/2$, where $z$ is the coordination number. 
Using the equivalence between the spin and lattice boson systems and the mean-field approximation to the spin system, Matsubara and Matsuda\,\cite{Matsubara} discussed the physical properties of the $\lambda$-transition in $^{4}$He. They showed that the BEC of the lattice bosons is equivalent to the magnetic ordering of the ferromagnet with $J^{\perp}\,{>}\,J^{\parallel}$. When the magnetic field $H$ is smaller than the saturation field $H_{\rm s}$, which is given by $H_{\rm s}\,{=}\,z(J^{\perp}\,{-}\,J^{\parallel})/2$ at $T\,{=}\,0$ K, the ordered moment is canted from the $xy$ plane, so that it has components both parallel and perpendicular to the magnetic field. This canted ferromagnetic state exactly corresponds to the Bose\,--\,Einstein condensed state of the lattice bosons. Bose-gas description of spin model was extended to Heisenberg antiferromagnets in high magnetic fields, and the phase transitions between fully polarized and antiferromagnetic ordered states or the spin structures of the ordered states were discussed from the BEC point of view \cite{Batyev,Nikuni2,Starykh}.

K$_2$CuF$_4$ is a well-known ferromagnetic insulator with a layered crystal structure closely related to the K$_2$NiF$_4$ structure\,\cite{Hirakawa2}. The crystal structure was first determined by Knox\,\cite{Knox} and later redetermined by Hidaka {\it et al.}\,\cite{Hidaka}.
In contrast to many other magnets with the K$_2$NiF$_4$ structure\,\cite{Legrand}, K$_2$CuF$_4$ has ferromagnetic exchange interactions and undergoes a ferromagnetic phase transition at $T_{\rm C}\,{=}\,6.25\,{\rm K}$\,\cite{YamadaI}. Owing to the antiferrodistortive arrangement of the elongated axes of CuF$_6$ octahedra, neighboring hole orbitals $d(x^2\,{-}\,y^2)$ are orthogonal to each other. This leads to ferromagnetic exchange interactions between neighboring spins in a square lattice layer parallel to the $ab$ plane\,\cite{Khomskii}. 

The magnetic model of K$_2$CuF$_4$ in a magnetic field parallel to the $c$ axis 
can be written as
\begin{eqnarray}
\mathcal{H}=\hspace{-1mm}&-&\hspace{-1mm}J\sum_{\langle i,j\rangle}{\bm S}_i\cdot {\bm S}_j+J_{\rm A}\sum_{\langle i,j \rangle}S^z_iS^z_j
-J^{\prime}\sum_{\langle l,m\rangle}{\bm S}_l\cdot {\bm S}_m\nonumber\\
\hspace{-1mm}&-&\hspace{-1mm}g\mu_{\rm B} H\sum_iS^z_i,
\label{eq:Hamiltonian}
\end{eqnarray}
where the first and second terms are the ferromagnetic exchange interaction and easy-plane anisotropy in the square lattice layer parallel to the $ab$ plane, respectively. The third term is the ferromagnetic exchange interaction between neighboring layers. The last term is the Zeeman term. The exchange parameters and $g$ factor were obtained as $J/k_{\rm B}\,{=}\,22.8\,\,{\rm K}$~\cite{Funahashi}, $J_{\rm A}/k_{\rm B}\,{=}\,0.22\,\,{\rm K}$~\cite{Kubo}, $J^{\prime}/k_{\rm B}\,{=}\,0.015-0.017\,\,{\rm K}$~\cite{Hirakawa3,Yamazaki} and $g\,{=}\,2.093$~\cite{Sasaki}. 
Because $J\,{\gg}\,J^{\prime}$ and $J_{\rm A}\,{>}\,0$, K$_2$CuF$_4$ is magnetically described as a spin-1/2, quasi-2D XXZ ferromagnet with weak anisotropy of the easy-plane type. Because the intralayer and interlayer exchange interactions are both ferromagnetic, there is no spin frustration between neighboring layers, as discussed for BaCuSi$_2$O$_6$\,\cite{Sebastian}. Therefore, the spin ordering in K$_2$CuF$_4$ for $H\,{\parallel}\,c$ is equivalent to the BEC of lattice bosons with $S^z\,{=}\,{+}\,1$, as discussed by Matsubara and Matsuda\,\cite{Matsubara}.

In this paper we present a phase diagram for temperature vs the magnetic field applied parallel to the $c$ axis in K$_2$CuF$_4$ and show that the phase boundary is described by the power law with exponent $\phi\,{\approx}\,1.0$ in a wide temperature range, as predicted by the theory\,\cite{Syromyatnikov}.

\section{Experimental details}
K$_{2}$CuF$_{4}$ crystals were prepared via the chemical reaction 2KF+CuF$_2 \rightarrow$ K$_2$CuF$_4$. KF and CuF$_2$ were dehydrated by heating in vacuum at about 100~${}^{\circ}$C.
The materials were then packed into a Pt tube. Single crystals were grown from the melt.
The temperature of the horizontal furnace was lowered from 850${}^{\circ}$C to 750~${}^{\circ}$C over 4 days. The crystals obtained were examined by X-ray powder diffraction and found to be K$_{2}$CuF$_{4}$. The crystals are easily cleaved parallel to the $c$ plane.

The magnetization was measured down to 1.8\,K in magnetic fields parallel to the $c$ axis using a SQUID magnetometer (Quantum Design MPMS-XL). A $^3$He system (iHelium3, IQUANTUM) was used for the measurement down to the lowest temperature of 0.5\,K.
The specific heat was measured down to 0.4 K by the relaxation method using a physical property measurement system (Quantum Design PPMS).

For the measurement of magnetization for $H\,{\parallel}\,c$, we used three samples, A, B and C, in the shape of rectangular plates with different thicknesses. Because the transition field was smaller than 3000 Oe, a correction for the demagnetizing field was necessary. The demagnetizing factor $N$ was calculated by a formula proposed by Osborn\,\cite{Osborn}. The internal field $H_{\rm int}$ is related to the external field $H_{\rm ext}$ as 
$H_{\rm int}\,{=}\,H_{\rm ext}\,{-}\,NM$,
where $M$ is the magnetization.

\section{Results and discussion}
\begin{figure}[t]
\begin{center}
\includegraphics[width=1.0\linewidth]{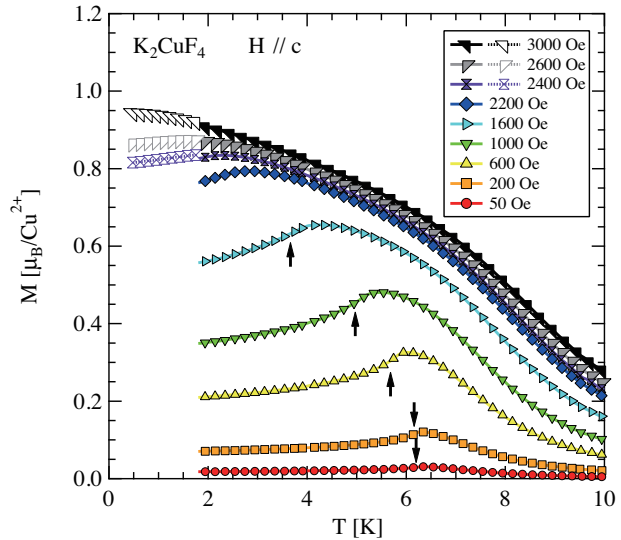}
\end{center}
\caption{(Color online) Temperature variation of magnetization in K$_2$CuF$_4$ measured at various external fields $H_{\rm ext}$ parallel to the $c$ axis. The data above and below 1.8 K were measured using samples A (or B) and C, respectively. Arrows indicate the transition temperature $T_{\rm C}$, which is assigned to the temperature with the peak in ${\rm d}M/{\rm d}T$.} 
\label{fig:MT}
\end{figure}
\begin{figure}[t]
\begin{center}
\includegraphics[width=1.0\linewidth]{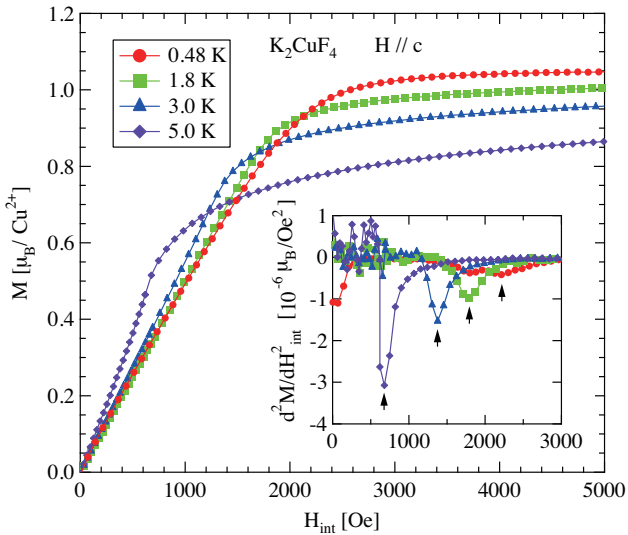}
\end{center}
\caption{(Color online) Magnetic-field dependence of the magnetization $M$ in K$_2$CuF$_4$ measured at various temperatures for $H\,{\parallel}\,c$, where the horizontal axis denotes the internal magnetic field $H_{\rm int}$. The inset shows its second field derivative $d^2M/dH_{\rm int}^2$ vs internal magnetic field $H_{\rm int}$. Arrows indicate the transition fields $H_{\rm c}(T)$.} \label{fig:MH}
\end{figure}

Figure~\ref{fig:MT} shows the temperature dependence of the magnetization in K$_2$CuF$_4$ measured at various external fields $H_{\rm ext}$ parallel to the $c$ axis. The magnetization data above and below 1.8 K were measured using samples A (or B) and C, respectively. The small discontinuous jump in the magnetization at 1.8 K for $H_{\rm ext}\,{\geq}\,2400$ Oe is ascribed to the difference in the internal field $H_{\rm int}$ due to the difference in the sample shape and, thus, the anomaly is extrinsic.

With decreasing temperature, the magnetization increases and shows a cusplike maximum. The temperature giving maximum magnetization decreases with increasing magnetic field. This magnetization behavior is consistent with that in a previous report\,\cite{Hirakawa}, where magnetization data down to 4.2 K were reported. In the molecular-field approximation for a 3D easy-plane-type ferromagnetic XXZ model, the magnetization increases up to the transition temperature as a convex function as the temperature decreases then becomes constant below $T_{\rm C}$\,\cite{Matsubara}. 
The existence of the cusplike maximum in the temperature dependence of the magnetization cannot be understood in terms of the molecular-field approximation. An advanced calculation will be necessary to describe the cusplike anomaly of magnetization from the approach of the spin system. This magnetization behavior is qualitatively described by the following lattice boson picture. In a lattice boson model\,\cite{Matsubara}, the density of bosons $\rho$ and ${\langle}\,{S^z}\,{\rangle}$ are related as ${\rho}\,{=}\,{\langle}\,{S^z}\,{\rangle}\,{+}\,1/2$. Because the boson density $\rho$ is larger than 1/2, it is convenient to consider holes instead of bosons, where the holes correspond to lattice points that are not occupied by bosons. The density of holes ${\rho}^{\prime}$ is related to ${\langle}\,{S^z}\,{\rangle}$ as ${\langle}\,{S^z}\,{\rangle}\,{=}\,1/2\,{-}\,{\rho}^{\prime}$. With decreasing temperature, the number of thermally excited holes decreases. At $T_{\rm C}$, the BEC of holes occurs and the number of condensed holes increases with decreasing temperature below $T_{\rm C}$ . The increase in the number of condensed holes exceeds the decrease in the number of thermally excited holes\,\cite{Nikuni,YamadaF1}. Consequently, the density of holes has a cusplike minimum at $T_{\rm C}$. This leads to a cusplike maximum of the density of lattice bosons, i.e., a cusplike maximum of the magnetization at $T_{\rm C}$.

Although the cusplike maximum of the magnetization at $T_{\rm C}$ can be derived by mean-field theory for 3D BEC, we assign the transition temperature $T_{\rm C}$ to the temperature of the peak in $dM/dT$ indicated by arrows in Fig.\,\ref{fig:MT}, because this temperature coincides with the temperature giving a peak in the specific heat below 200 Oe, where $T_{\rm C}$ has little dependence on the internal magnetic field. 

Figure~\ref{fig:MH} shows the magnetization curves for $H\,{\parallel}\,c$ measured at various temperatures for K$_2$CuF$_4$. The horizontal axis is the internal magnetic field $H_{\rm int}$ calculated using the magnetization and the demagnetizing factor. The inset shows its second derivative 
$d^2M/dH_{\rm int}^2$ vs $H_{\rm int}$. 
In a weak field, the magnetization is proportional to the internal field and its slope becomes smaller with decreasing temperature, which is consistent with the temperature dependence of the magnetization shown in Fig.\,\ref{fig:MT}. In a strong field, the magnetization curve behaves similarly to the Brillouin function. Here, we assign the transition temperature $H_{\rm c}$ to the field with the cusplike minimum in $d^2M/dH_{\rm int}^2$, indicated by arrows in the inset of Fig.\,\ref{fig:MH}, because the transition points are consistent with those obtained from the temperature dependence of the magnetization shown by arrows in Fig.\,\ref{fig:MT}.

\begin{figure}[t]
\begin{center}
\includegraphics[width=1.0\linewidth]{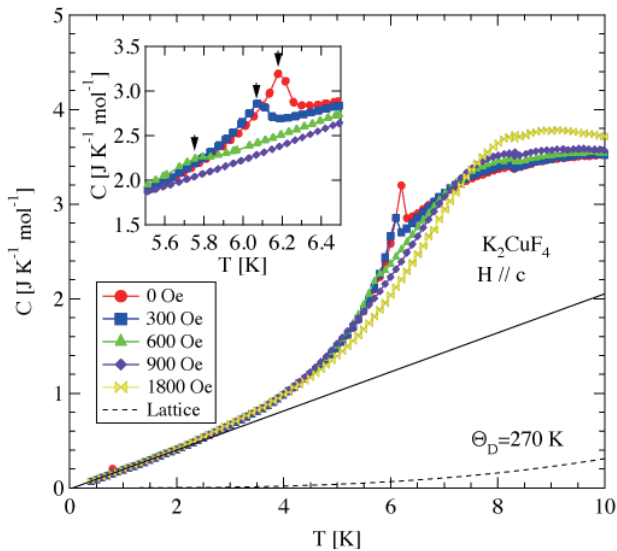}
\end{center}
\caption{(Color online) Temperature variation of total specific heat in K$_2$CuF$_4$ measured at various external magnetic fields parallel to the $c$ axis. The dashed line denotes the lattice contribution estimated from the Debye $T^3$-law with a Debye temperature of $\Theta_{\rm D}\,{=}\,270$~K\,\cite{YamadaI}. The solid line denotes the fit using Eq.\,(\ref{eq:takahashi}) with $J/k_{\rm B}\,{=}\,21.0$ K. The inset shows a magnification of the curves around 6 K, where a $\lambda$-like anomaly associated with the ferromagnetic phase transition is observed. Arrows indicate transition temperatures.} 
\label{fig:CT}
\end{figure}

\begin{figure}[t]
\begin{center}
\includegraphics[width=1.0\linewidth]{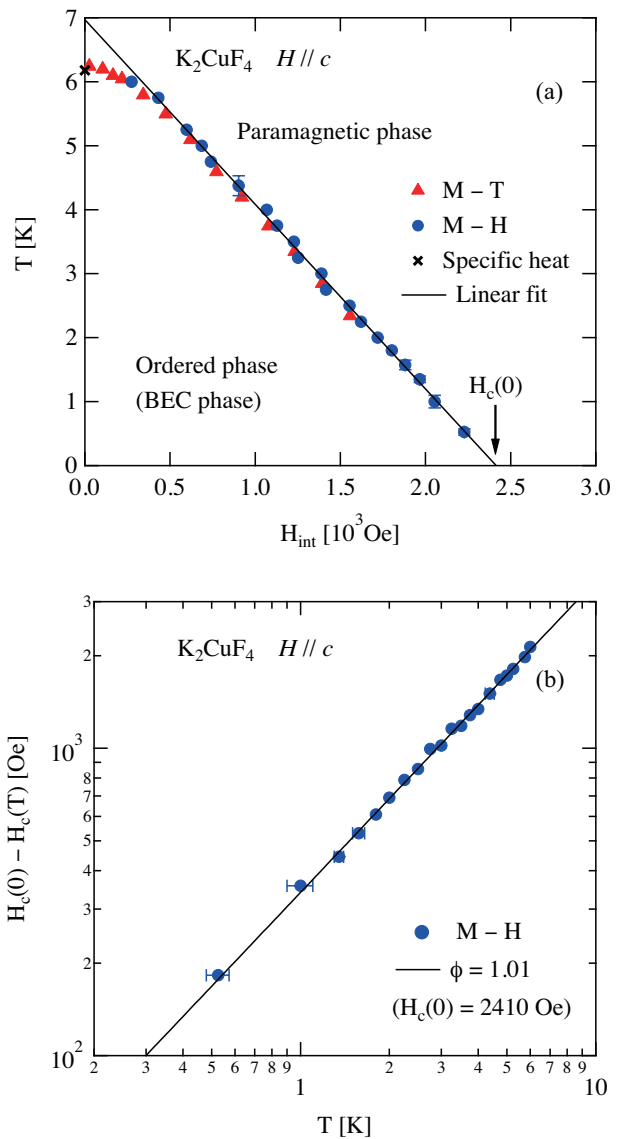}
\end{center}
\caption{(Color online) (a) Temperature vs internal magnetic field phase diagram for K$_2$CuF$_4$ with $H\,{\parallel}\,c$. Closed triangles and circles are transition points determined from the temperature and field dependences of the magnetization, respectively. The cross mark ($\times$) is the Curie temperature obtained by specific heat measurement at zero magnetic field. The solid line is the linear fit to the phase boundary below 5 K. (b) Double logarithmic plot of $H(T)-H_{\rm c}(0)$ against $T$ with $H_{\rm c}(0)\,{=}\,2410$ Oe.
The solid line is a fit with $\phi$\,{=}\,1.01.} 
\label{fig:Phase}
\end{figure}

Figure~\ref{fig:CT} shows the low-temperature specific heat measured at various external magnetic fields. The dashed line denotes the lattice contribution estimated from the Debye $T^3$-law with a Debye temperature of $\Theta_{\rm D}\,{=}\,270$~K\,\cite{YamadaI}. A small discontinuous anomaly at approximately 9 K arises from an instrumental problem and is not intrinsic. Below 3 K, where the lattice contribution is negligible, the specific heat is proportional to the temperature $T$, which is a characteristic of 2D Heisenberg ferromagnets\,\cite{Takahashi}. This confirms that K$_2$CuF$_4$ has good two-dimensionality.
The sharp $\lambda$-like anomaly at approximately 6~K indicates a magnetic phase transition.
As shown in the inset, the $\lambda$-like anomaly becomes small as the external magnetic field increases. Above 900 Oe, the transition temperature is undistinguishable. 

From the spin-wave theory, the low-temperature magnetic specific heat $C_{\rm M}$ of a 2D Heisenberg ferromagnet is given as\,\cite{Takahashi}
\begin{eqnarray}
C_{\rm M}=\frac{\pi R}{12S}\frac{k_{\rm B}}{J}T.
\label{eq:takahashi}
\end{eqnarray}
Fitting the experimental specific heat below 2.0 K, we obtain $J/k_{\rm B}\,{=}\,21.0\,{\pm}\,0.4$~K.
This value of $J$ is somewhat larger than $J/k_{\rm B}\,{=}\,17.6$~K obtained by Yamada\,\cite{YamadaI}, who used the specific heat data between 1.3 and 3 K, but is smaller than $J/k_{\rm B}\,{=}\,22.8$~K obtained from the dispersion relation by Funahashi {\it et al}.\,\cite{Funahashi}.


Figure~\ref{fig:Phase}(a) shows the phase diagram for temperature vs internal magnetic field for K$_2$CuF$_4$ with $H\,{\parallel}\,c$. The transition points determined from the temperature and magnetic field dependences of the magnetization are consistent with each other. We can see that the phase boundary is linear in a wide temperature range below 5 K in accordance with the theory\,\cite{Syromyatnikov}. The solid line in Fig.\,\ref{fig:Phase}(a) is a linear fit to the experimental phase boundary below 5 K. From the linear fit, we obtain the transition field $H_{\rm c}(0)$ at $T\,{=}\,0$ K to be $H_{\rm c}(0)\,{=}\,2415$ Oe. We also apply the power law in Eq.~(\ref{eq:power}) to the experimental phase boundary between 0.53 and 5 K. The best fit is obtained for ${\phi}\,{=}\,1.01$ and $H_{\rm c}(0)\,{=}\,2410$ Oe. Figure~\ref{fig:Phase}(b) shows a double logarithmic plot of $H(T)-H_{\rm c}(0)$ against $T$ with $H_{\rm c}(0)\,{=}\,2410$ Oe. The solid line is the linear fit with ${\phi}\,{=}\,1.01$. From these results, we deduce that quasi-2D BEC occurs in this temperature range. Because the interlayer exchange interaction in K$_2$CuF$_4$ is $J^{\prime}/k_{\rm B}\,{=}\,0.015-0.017$~K~\cite{Hirakawa3,Yamazaki}, the crossover from quasi-2D BEC to 3D BEC described by setting ${\phi}\,{=}\,3/2$ is expected to occur at approximately 20 mK, which is much lower than the present temperature range. 

\section{Summary}
In conclusion, we have presented the results of magnetization and specific heat measurements on the $S\,{=}\,1/2$, quasi-2D, easy-plane-type XXZ ferromagnet K$_2$CuF$_4$ for $H\,{\parallel}\,c$, which is equivalent to the lattice boson model described by Eq.~(\ref{eq:Lattice_boson}). We obtained the phase diagram for temperature vs internal field as shown in Fig.~\ref{fig:Phase}(a). The phase boundary between the polarized paramagnetic and ordered phases is described by the power law in Eq.~(\ref{eq:power}) with exponent ${\phi}\,{\approx}\,1.0$ in a wide temperature range below 5 K. This result is in agreement with the theory of quasi-2D BEC universality\,\cite{Syromyatnikov}.

\section*{ACKNOWLEDGMENT}
This work was supported by Grants-in-Aid for Scientific Research (A) (Nos. 23244072 and 26247058) and (C) (No. 16K05414), and a Grant-in-Aid for Young Scientists (B) (No. 26800181) from Japan Society for the Promotion of Science.

\end{document}